\documentclass[12pt, a4paper]{article}
\usepackage{a4}

\usepackage{graphicx}
\usepackage{amsmath,amssymb,latexsym,mathrsfs}

\newcommand{\onef}{\mbox{$1_f$}}

\def\ppall{\mathaccent23p}

\begin{document}

%%\draft  % \draft command makes pacs numbers print

\vspace{20mm}
\begin{center}
{\LARGE \bf 
Twisted mass QCD and the FNAL heavy quark formalism.
}\\[15mm] 

{\bf C.McNeile \\
Department of Physics and Astronomy, The Kelvin Building
University of Glasgow, Glasgow G12 8QQ, U.K.
}\\[2mm]
 
\end{center}

\begin{abstract}
At tree level, I discuss modifying the FNAL heavy quark formalism to
include a twisted mass term. I find that at maximal twist the 
so called KLM factor is independent of the heavy mass.
 \end{abstract}
% Lattice QCD Calculations
%%\pacs{12.38.Gc} 

% body of paper here

\section{Introduction and motivation}

Although only recently developed, twisted mass QCD
is already proving to be an excellent technique for
producing accurate lattice QCD results.
Twisted mass lattice 
QCD~\cite{Frezzotti:2000nk,Frezzotti:2003ni} calculations have
been used to test and 
constrain chiral perturbation 
theory~\cite{Boucaud:2007uk,Blossier:2007vv,Boucaud:2008xu}, study
baryons~\cite{Alexandrou:2008tn}, 
heavy-light mesons~\cite{Blossier:2007pt}, 
flavour singlet mesons~\cite{Jansen:2008wv},
static-light mesons~\cite{Jansen:2008si},
and the pion form factor~\cite{Simula:2007fa}. 
A key theoretical
advantage of the twisted mass formalism is that the action is
automatically $O(a)$ improved
at maximal twist~\cite{Frezzotti:2003ni,Frezzotti:2005gi}.
Twisted mass QCD has recently been 
reviewed~\cite{Shindler:2007vp,Sint:2007ug}.

%%
%%  heavy quark mass effects power counting
%%
An important part of the twisted mass program is that
there are power counting arguments to understand the $O(a^2)$
corrections for light 
quarks~\cite{Frezzotti:2007qv}. 
These are based on a Symannzik analysis that is suitable for
light quarks. 
As the mass of the heavy quark increases 
towards the mass of the charm quark and beyond,
then
$O((a M_Q)^2)$ effects may become sizable. One way to estimate
heavy quark mass effects is to use the FNAL heavy quark mass
formalism~\cite{ElKhadra:1996mp}.

The full FNAL heavy quark formalism requires tuning the terms
in the  heavy quark action. For example the  clover coefficient
of the clover term and the coefficient of the spatial Wilson 
term.
There have been a few numerical 
studies of the required
tuning~\cite{ElKhadra:1997hq,Sroczynski:1999he,Sroczynski:2000hg,Lin:2006ur,Christ:2006us,Kayaba:2006cg}.
However, many groups have used the FNAL formulation
to estimate heavy mass corrections to 
decay constants and 
masses~\cite{Bernard:1993zh,Bernard:1998xi,Bernard:2002pc}
for the standard Wilson and clover actions. The prescription was to 
use the kinetic mass and to
multiply
the quark fields by the KLM factor in equation~\ref{eq:KLMsimple}
\begin{equation}
Z_{KLM} = \sqrt{2 \kappa e^{m_0} }
\label{eq:KLMsimple}
\end{equation}
where $m_0$ is defined in
\begin{equation}
m_0 = \log ( \frac{1}{2 \kappa} - 3 ).
\end{equation}
and $\kappa$ is the standard hopping parameter used in the clover and Wilson
actions. The corrections to equation~\ref{eq:KLMsimple} that include
non-perturbative O(a) mass corrections 
are in~\cite{Becirevic:1998ua}.

A numerical test of the KLM factors is reported by 
El-Khadra et al.~\cite{ElKhadra:1997hq}
and a test of the FNAL formalism for the charm 
mass reported by Dougall et al.~\cite{Dougall:2005ev}.
A critical comparison of the KLM factors 
to renormalisation factors determined non-perturbatively 
is in~\cite{Crisafulli:1997ic}. 
The aim of this paper is to find the equivalent KLM
factor for twisted mass fermions at tree level.

Throughout this paper I will mostly only consider tree level
perturbation theory.  This is in the spirit of estimating and
correcting the leading $O((a m_Q)^n)$ corrections to the results
of twisted mass calculations.
I also don't consider the two doublet 
twisted mass formalism for
including non-degenerate quarks in unquenched
calculations~\cite{Frezzotti:2003xj,Chiarappa:2006ae}.
The results will be useful to analyse existing
$n_f$=2 twisted mass calculations and twisted mass
calculations that use the 
Osterwalder-Seiler action~\cite{Osterwalder:1977pc,Frezzotti:2004wz}
for the heavy quarks in unquenched calculations with 
$2+1+1$ flavours of sea quarks.

The theoretical foundations of twisted mass
QCD~\cite{Frezzotti:2000nk,Frezzotti:2001ea,Frezzotti:2003ni} 
use a mass independent renormalisation
scheme~\cite{Weinberg:1951ss}, but the FNAL heavy quark formalism
uses mass dependent renormalisation factors. The issue of mass
dependent versus mass independent renormalisation schemes
is reviewed by Georgi~\cite{Georgi:1994qn} and
Kronfeld~\cite{Kronfeld:2002pi}.

\section{A brief introduction to twisted mass QCD}

I first review the twisted mass quark action in the continuum.
\begin{equation}
S_F  = \int d^4 x \overline{\chi} ( \gamma_\mu D_\mu + m_q + i \mu_q \gamma_5 \tau^3) \chi
\label{eq:conttwistbasis}
\end{equation}
where $\tau^3$ is the third Pauli spin matrix in flavour space, and 
$m_q$ and $\mu_q$ are mass parameters. The fields $\chi$ 
and $\overline{\chi}$ are in the twisted basis. The quarks fields
can be transfered to what is known as the "physical basis"
by the transformation:
\begin{equation}
\psi = exp( i \omega \gamma_5 \tau^3 / 2 ) \chi \; \; \;
\overline{\psi} = \overline{\chi} exp( i \omega \gamma_5 \tau^3 / 2 ) 
\label{eq:trans}
\end{equation}
where $\tan \omega = \mu_q / m_q$.  After the transformation in 
equation~\ref{eq:trans}, the twisted quark action in the physical 
basis is
\begin{equation}
S_F  = \int d^4 x \overline{\psi} ( \gamma_\mu D_\mu + \sqrt{m_q^2 + \mu_q^2} ) \psi
\end{equation}

The lattice version of the twisted mass 
action in equation~\ref{eq:conttwistbasis} 
is written down in the standard way. 
However, to prepare for applying the FNAL heavy quark formalism,
I consider a twisted
version ($S_{lat,F}$) of
the action written down by Lin and Christ~\cite{Lin:2006ur}.
Lin and Christ~\cite{Lin:2006ur} have a table
that summaries the choices of coefficients used by other
heavy quark formulations~\cite{ElKhadra:1996mp,Aoki:2001ra}. 
The version of the twisted Wilson action used in existing 
numerical calculations
uses $\zeta$ = $r_s$ =1. I only use arbitary value of
$\zeta$ and $r_s$ in section~\ref{se:auto}.
The twisted QCD action in the twisted basis is
\begin{equation}
S_{lat,F} = \sum_{x} \overline{\chi}(x)  
( \gamma_0 D_0 + \zeta \gamma_i D_i
 - \frac{1}{2} D_0^2 - r_s \frac{1}{2}D_i^2
+ m_{q} +  i \tau_3 \gamma_5 \mu_q )  \chi(x)
\label{eq:twistchrist}
\end{equation}
%%%
The derivatives are defined by
\begin{equation}
D_\mu \psi(x) = \frac{1}{2} 
[ U_\mu(x) \psi(x+\hat{\mu}) - 
 U_\mu^\dagger(x) \psi(x - \hat{\mu})] 
\end{equation}
%%%
and 
\begin{equation}
D_\mu^2 \psi(x) = 
U_\mu(x) \psi(x+\hat{\mu}) + 
 U_\mu^\dagger(x) \psi(x - \hat{\mu})
- 2 \psi(x) 
\end{equation}
%%%
%%
%%
The special choice of  $\omega = \pi / 2$ is known 
as maximal twist, where all of the quark mass is in the 
$\gamma_5 \tau^3$ term.
At maximal twist there are no
$O(a)$ corrections to the continuum 
result~\cite{Frezzotti:2003ni,Frezzotti:2005gi}.
Achieving maximal twist is 
non-trivial, because
of the additive mass renormalisation of the Wilson formulation.
but achievable numerically in practise.

\section{The FNAL heavy quark formalism for Wilson fermions}

The FNAL formalism for heavy quarks was originally described 
by~\cite{ElKhadra:1996mp}. Further 
developments of the FNAL lattice heavy quark formulation
are described 
in~\cite{Aoki:2001ra,Kronfeld:2002pi,Lin:2006ur,Christ:2006us,Oktay:2008ex}
Here I review 
the free field calculation from El-Khadra et
al.~\cite{ElKhadra:1996mp} as warm up
to adding a twisted mass term.

The formalism starts with the quark propagator in momentum and time.
\begin{equation}
S(t,\vec{p}) =   e^{ - E  t} 
\frac{
\sinh E \gamma_0 \mbox{sign}(t) 
- i  \gamma_i  \ppall_i + m_q + 1 - \cosh E + \frac{1}{2}
\hat{p}_i \hat{p}_i
}
{
2 Z_2 \sinh E
}
\label{eq:quarwilson}
\end{equation}
%%%
%%
%%
where $\ppall_i = \sin ( a p_i) $ and  
$\hat{p}_i = 2 \sin ( \frac{a p_i} {2}) $.
%%%
and $Z_2$ is calculated to be
\begin{equation}
Z_2 = 1 + m_q a + \frac{1}{2} \hat{p}^2 a^2 
\end{equation}
Equation~\ref{eq:quarwilson} is only valid 
for $t >0$, because there
is an additional term at $t=0$. As $a m_q$ gets very large,
$1/Z_2$ will get very small and this will cause problems
with the dynamics.

In the FNAL formulation, when the quark mass gets heavy,
the dispersion relation of the heavy quark gets modified to
\begin{equation}
E^2 = M_1^2 + \frac{M_1}{M_2} \vec{p}^2 + ...
\label{eq:disp}
\end{equation}
%%%
where $M_1$ is known as the rest mass and
$M_2$ is called the "kinetic" mass. Another closely
related way to measure the deviations of the lattice dispersion
relation from the continuum one is via 
the ``speed of light''~\cite{Alford:1996nx}.

\begin{equation}
M_1 = \frac{1}{a} \ln ( 1 + m_q a )
\end{equation}
Expanding the rest mass in terms of $m_q a $ gives
%%%%
\begin{equation}
M_1 = m_q - \frac{1}{2} m_q^2 + \frac{1}{3} m_q^3 + O( m_q^4) 
\label{eq:wilsonEXpand}
\end{equation}
%%%
The second term in the expansion is the leading $b_m$
improvement term in the 
ALPHA formulation of the clover action~\cite{Sint:1997jx}. 
The connection between
the ALPHA formulation of the clover fermion action with the FNAL
formulation of the heavy fermion action is 
demonstrated at one loop by Mertens et al.~\cite{Mertens:1997wx}
for the quark mass.

The $M_2$ kinetic mass is extracted 
using~\cite{ElKhadra:1996mp} 
%%%
\begin{equation}
\frac{1}{M_2} = \frac{\partial^2 E}{\partial p_1 \partial p_1 }
\mid_{p_1=0}
\label{eq:kinmassDEFN}
\end{equation}
%%%
For Wilson fermions the standard result is
\begin{equation}
\frac{1}{M_2} = 
\frac{2}{m_q a (2 + m_q a  ) }
+
\frac{1}{1 + m_q a }
\end{equation}
%%%
with an expansion in the quark mass:
\begin{equation}
M_2 = a m_q - \frac{1}{2} ( a m_q )^2 + 
 ( a m_q )^3 - \frac{7}{4}  ( a m_q )^4
+ O( ( a m_q )^5  )
\end{equation}

The KLM factor is defined~\cite{Bernard:1993zh}
via equation~\ref{eq:KLMdefn}.
\begin{equation}
Z_{KLM}^2 \sum_x 
\langle 0 \mid 
\psi(x) \overline{\psi} (0) 
\mid 0 \rangle^{\mbox{latt}}
=
\int d^3 x
\langle 0 \mid 
\psi(x) \overline{\psi} (0) 
\mid 0 \rangle^{\mbox{cont}}
\label{eq:KLMdefn}
\end{equation}
hence
\begin{equation}
Z_{KLM} = \sqrt{Z_2 (\vec{p}=0)} = \sqrt{1 + m_q}
\end{equation}
There is a prescription for constructing amplitude factors
for operators that are extended 
in space and time, such as conserved 
currents~\cite{Bernard:1993fq,Crisafulli:1997ic}
that I do not discuss.

\section{The FNAL heavy quark formalism for twisted mass QCD}

For the twisted mass formulation I first match the twisted heavy
action onto a twisted continuum fermion 
action in equation~\ref{eq:trans}, then rotate back to
the standard continuum action in the physical basis. This two step
procedure seems more natural than trying to match the action in the 
twisted basis back to the continuum Dirac action in one step.
I add the superscript $T$ to show that the quantities are for the 
twisted action.

The quark propagator in time and spatial momentum 
for Wilson twisted mass fermions
has been written down by 
Cichy et al.~\cite{Cichy:2008gk} in the twisted basis,
as part of their study of 
the pion and nucleon correlators in free field theory
for the twisted mass action and a variety of actions that 
obeyed the Ginsparg-Wilson relation. 
\begin{eqnarray}
S(\vec{p},t) & = &\frac{1} {2 Z^T_2 \sinh E^T} ( 
\onef ( \mbox{sgn}(t) \sinh E^T \gamma_4 -  i  \gamma_i \ppall_i ) 
 \nonumber \\
& + & [ ( 1 - \cosh E) + a m_q +   \frac{1}{2}   \hat{p}_i\hat{p}_i  ] ) 
- i a \mu_q \gamma_5 \tau^3 ) e^{ - E^T t}
\end{eqnarray}

The rest mass ($M_1$) is obtained~\cite{Frezzotti:2001ea,Shindler:2007vp}
from the energy ($E^T$) at zero three
momentum:
\begin{equation}
\cosh M_1^T =1 + 
\frac{ a^2 m_q^2 + a^2 \mu_q^2 }
{2 (1 + a m_q ) }
\label{eq:twistpole}
\end{equation}
%%%
At maximal twist, equation~\ref{eq:twistpole} shows the 
pole mass $M_1^T$
is a function of the $(a \mu_q)^2$ so there will be no dependence
on odd powers of the lattice spacing, consistent with the general
symmetry arguments~\cite{Shindler:2007vp}.
The corrections to the continuum $M_1$ = $a \mu_q$ are
much smaller than for the Wilson action
in equation~\ref{eq:wilsonEXpand}.
\begin{equation}
M_1^T = a \mu_q - \frac{1}{24} (a \mu_q)^3 + \frac{3}{640} (a \mu_q)^5 + O( (a \mu_q)^7 ) 
\label{eq:twistpoleperturb}
\end{equation}

The kinetic mass ($M_2^T$) is 
\begin{equation}
M_2^T =  a \mu_q \frac{\sqrt{4 + (a \mu_q)^2}} {2 - (a\mu_q)^2  } 
\end{equation}
The expansion of $M_2^T$ in terms of $\mu_q$ 
%%%
\begin{equation}
M_2^T =  a \mu_q + \frac{5}{8} (a \mu_q)^3 + \frac{39 }{128} (a \mu_q)^5
+ O( (a \mu_q)^7 )
\end{equation}

The mass dependent amplitude is 
%%%
\begin{equation}
Z_{KLM}^T = \sqrt{Z^T_2(\vec{p}=0)}  =  \sqrt{1 + m_q}
\label{eq:twistamp}
\end{equation}
At maximal twist $Z_{KLM}^T$ is 1,
because $m_q$ is tuned to zero.
Is is surprising that $Z_{KLM}^T$ is independent of
$\mu_q$,
because I would have naively expected an expression that
depended on the twisted mass $\mu_q$, but with no $O(a)$ errors.
In the calculation 
$Z_{KLM}$ is independent of $\mu_q$, because as
Shindler~\cite{Shindler:2007vp} notes, the $\mu_q$ term and the Wilson terms
"point in different directions" so don't interfere. However it
would be "cooler" to have a deeper more theoretical argument.
The results for the $M_1$ and $M_2$ masses do
show a dependence on $\mu a$, beyond the continuum result,
but with no $O(a)$ terms
as expected.

The KLM factor was originally obtained as part of
deriving the transfer matrix for Wilson 
fermions~\cite{Luscher:1976ms,ElKhadra:1996mp}. 
The derivation of the transfer matrix
was extended to twisted mass QCD by 
Frezzotti et al.~\cite{Frezzotti:2001ea,Shindler:2007vp}.
The normalisation of the fields in the derivation of 
the transfer matrix depended on a matrix called "B"
in equation 13 in~\cite{Luscher:1976ms}. The equvalent
"B" matrix for twisted mass QCD is the same as for
Wilson fermions and independent of the $\mu_q$ 
mass~\cite{Frezzotti:2001ea,Shindler:2007vp}, and so
is consistent with the KLM factor being independent
of $\mu_q$ at tree level.
Although this analysis is focused towards twisted
mass fermions, it is interesting to try and 
understand the $\mu_q$ independence of the KLM
factor. One way of getting some insight it to look
at lattice actions with more symmetry such as those
with Ginsparg-Wilson 
symmetry~\cite{Narayanan:1992wx,Neuberger:1997fp,Neuberger:1998wv}, 
or those with a 
remnant of chiral symmetry such as improved staggered 
actions~\cite{Follana:2006rc}.

% overlap
Liu and Dong have studied $O((a m_Q)^2)$ 
and $O((a^2 m_Q \Lambda_{QCD})$ 
effects in renormalisation
constants and the dispersion relation
in numerical data~\cite{Liu:2002qu,Dong:2007da}. 
They found that the variant of the overlap action they
used had lattice errors under control if they
kept $a m_Q < 0.5$. This numerical work suggests that
a KLM factor for overlap fermions does depend on the 
heavy quark mass, although there are no $O(a m_Q)$ corrections
as expected.
The work by the TWQCD collaboration uses
the overlap action with much larger masses to study
mesons containing the bottom quark~\cite{Chiu:2007km}.

Aarts and Foley~\cite{Aarts:2006em} have studied an 
overlap
operator~\cite{Neuberger:1997fp,Neuberger:1998wv}
in free field theory.
They~\cite{Aarts:2006em} find an overall
mass dependent renormalisation factor for the quark propagator
that suggests a mass dependent KLM factor.
However, there are a wide variety of different solutions to the 
Ginsparg-Wilson relation, some of which will have a different
mass dependence. Liu and Dong~\cite{Liu:2002qu} discuss
one choice that may have good properties in the heavy mass
limit.
It would be interesting to see if a KLM factor could 
parameterise the numerical data of~\cite{Liu:2002qu,Dong:2007da}
using methods in~\cite{Aarts:2006em,Li:2006kk}.

%%
%% HISQ
%%
The tree level mass corrections to the improved staggered
action called HISQ were considered by the HPQCD 
collaboration~\cite{Follana:2006rc}. The coefficient of 
the Naik term was 
tuned at tree level to obtain a speed of light
of one,
up to errors of order $O((a m_q)^{12})$.
%%%
The wave function renormalisation at tree level reported by
HPQCD~\cite{Follana:2006rc} for the HISQ action with the Naik term
corrected with a mass dependent factor, had an explicit but weak
dependence on the quark mass.

From considering the HISQ and overlap actions above, 
it is unusual to have a wave function
factor that does not depend on the physical quark mass.
\section{Automatic $O(a)$ improvement and the FNAL formalism}
\label{se:auto}

One very clever proof for the automatic $O(a)$ 
improvement~\cite{Frezzotti:2005gi}, used 
that the action in the physical basis 
($S_{lat,F}$ in equation~\ref{eq:christ} with 
 $\zeta$=1 and $r_s$=1)
was invariant under
the symmetry $ {\cal P} \times {\cal D}_d \times ( \mu_q \rightarrow -\mu_q)  $
where ($x_P = (-x ,t)$). The automatic $O(a)$ 
improvement of the 
twisted mass Wilson action,
based on the above symmetry,
has been tested numerically in 
quenched 
QCD~\cite{DellaMorte:2001ys,Jansen:2005gf,Jansen:2003ir} and
$n_f = 2$ unquenched QCD~\cite{Frezzotti:2007qv}.

The ${\cal P}$ symmetry transformation is defined
by
\begin{eqnarray}
U_0(x)  & \rightarrow & U_0(x_P)  \nonumber \\
U_k(x)  & \rightarrow & U_k^\dagger(x_P - a \hat{k})  \nonumber \\
\psi(x) & \rightarrow & \gamma_0 \psi( x_P ) \nonumber \\
\overline{\psi}(x) & \rightarrow &  \overline{\psi} ( x_P ) \gamma_0 \nonumber
\end{eqnarray}
and the ${\cal D}_d $ symmetry is defined by
%%%%
\begin{eqnarray}
U_\mu(x) & \rightarrow & U^\dagger_\mu( - x - a \hat{\mu}) \nonumber \\
\psi(x)  & \rightarrow &  e^{ 3 i \pi / 2 } \psi(-x)  \nonumber \\
\overline{\psi}(x)  & \rightarrow &  e^{ 3 i \pi / 2 } \overline{\psi}(-x)  \nonumber
\end{eqnarray}

The argument in~\cite{Frezzotti:2005gi} only required that the 
action was invariant under the group 
 $ {\cal P} \times {\cal D}_d \times ( \mu_q \rightarrow - \mu_q)  $
and not that the action is also invariant under the full hypercubic
group.
To connect with the proof in~\cite{Frezzotti:2005gi},
I consider the 
twisted mass action in equation~\ref{eq:twistchrist}
at maximal twist rotated into the 
physical basis in equation~\ref{eq:christ}
%%%
\begin{equation}
S_{lat,F} = \sum_{x} \overline{\psi}(x)  ( \gamma_0 D_0 + 
\zeta \gamma_i D_i + \mu_q
 -  i \tau_3 \gamma_5 ( - \frac{1}{2} D_0^2 - r_s \frac{1}{2}D_i^2
+ m_{cr} )  \psi(x)
\label{eq:christ}
\end{equation}
%%%
where $m_{cr}$ is $m_q$ tuned to the critical mass from setting the
PCAC mass to zero.

The action $S_{lat,F}$ in 
equation~\ref{eq:christ}, with 
arbitrary $\zeta$ and $r_s$ parameters
 is also invariant under
 $ {\cal P} \times {\cal D}_d \times ( \mu_q \rightarrow -\mu_q)  $,
hence it should be automatically $O(a)$ improved. The clover term
should have a coefficient that is an odd power of the quark mass.
The two parameters: $\zeta$ and $r_s$,
need to be tuned for the twisted version of the 
heavy quark action, but should only be an even power of the quark mass.
The recent work on an improved Fermilab heavy quark action included
dimension 7 operators~\cite{Oktay:2008ex}, so adding a twisted mass
term, may help with the design of more highly improved heavy quark actions.

I don't see any simple connection between the symmetry
$ {\cal P} \times {\cal D}_d \times ( \mu_q \rightarrow -\mu_q)  $ and
getting a mass independent KLM factor in equation~\ref{eq:twistamp}.

\section{Conclusions}

I have discussed the inclusion of a twisted mass term with the 
FNAL heavy fermion action at tree level. This is
useful for the analysis of existing $n_f=2$ twisted mass
lattice QCD calculations with heavy masses,
and heavy quark calculations
using the Osterwalder-Seiler 
action~\cite{Osterwalder:1977pc,Frezzotti:2004wz}
on configurations with 2+1+1 flavours of sea quarks.
One surprising thing about the KLM factor for twisted
mass QCD was that it was independent of the twisted mass
at tree level.
To estimate the order of magnitude of the various improvements
terms I use the numerical values 
$\alpha_s \sim 0.24$  and $a m_Q \sim 0.26 $ for the 
$\beta = 3.9$ data set, with a lattice spacing of 0.0855 fm, from the 
ETM Collaboration~\cite{Blossier:2007pt}.
For the twisted quark action with heavy mass $m_Q$, automatic
$O(a)$ improvement means that the leading error should
be $O((a m_Q)^2)$ which is approximately 7\%. The mass independence
of the KLM factor for twisted mass fermions implies
that the leading corrections are $O((\alpha_s a m_Q)^2)$
and numerically about 2\%. The preliminary numerical results from lattice
QCD calculations with heavy quark from twisted mas QCD
seem to have larger errors than the above 
estimates~\cite{Dimopoulos:2008ee}.

I showed that the symmetry that protects the twisted Wilson
action from $O(a)$ corrections, should also protect 
an action where the hyper-cubic invariance is broken,
as used in the FNAL heavy quark action. Some quantities
such as the hyperfine spitting in charmonium are known
to be sensitive to the value of the clover coefficient
and lattice spacing 
errors~\cite{Choe:2003wx,diPierro:2003bu,McNeile:2004wu,Detar:2007ni} 
so automatic $O(a)$ improvement should be useful.

Since twisted mass QCD has no $O(a)$ errors, it is in principle
possible to use lattice calculations at three different lattice
spacings and take a consistent continuum limit for calculations that
include heavy quarks. 
It may be useful to supplement the
"brute force approach" with an estimate of systematic errors from the
FNAL heavy quark formalism.

\section{Acknowledgements}

I thank Chris Michael, Andrea Shindler,
and Christine Davies for discussions.

%%\bibliographystyle{h-physrev2}
%%\bibliography{heavytwist} 

\end{document}